\documentclass[preprint,trackchanges]{aastex6} %
\usepackage{bm, amsmath}

\newcommand{\pd}[2]{\displaystyle \frac{\partial #1}{\partial #2}}
\newcommand{\edy}[1]{#1}
\newcommand{\elm}[1]{}

\newcommand*\patchAmsMathEnvironmentForLineno[1]{%
  \expandafter\let\csname old#1\expandafter\endcsname\csname #1\endcsname
  \expandafter\let\csname oldend#1\expandafter\endcsname\csname end#1\endcsname
  \renewenvironment{#1}%
     {\linenomath\csname old#1\endcsname}%
     {\csname oldend#1\endcsname\endlinenomath}}%
\newcommand*\patchBothAmsMathEnvironmentsForLineno[1]{%
  \patchAmsMathEnvironmentForLineno{#1}%
  \patchAmsMathEnvironmentForLineno{#1*}}%
\AtBeginDocument{%
\patchBothAmsMathEnvironmentsForLineno{equation}%
\patchBothAmsMathEnvironmentsForLineno{align}%
\patchBothAmsMathEnvironmentsForLineno{flalign}%
\patchBothAmsMathEnvironmentsForLineno{alignat}%
\patchBothAmsMathEnvironmentsForLineno{gather}%
\patchBothAmsMathEnvironmentsForLineno{multline}%
}

\begin{document}

\title{Effects of Chemistry on Vertical Dust Motion in Early Protoplanetary Disks}
\author{Yoshinori Miyazaki \altaffilmark{1} and Jun Korenaga \altaffilmark{1}}
\affil{Department of Geology and Geophysics, Yale University, New Haven, Connecticut, USA}

\submitted{Submitted for Publication in The Astrophysical Journal}

\begin{abstract}
We propose the possibility of a new phenomenon affecting the settling of dust grains at the terrestrial region in early protoplanetary disks. Sinking dust grains evaporate in a hot inner region during the early stage of disk evolution, and the effects of condensation and evaporation on vertical dust settling can be significant. A 1-D dust settling model considering both physical and chemical aspects is presented in this paper. Modeling results show that dust grains evaporate as they descend into the hotter interior and form a ``condensation front,'' above which dust-composing major elements, Mg, Si, and Fe, accumulate, creating a large temperature gradient. Repeated evaporation at the front inhibits grain growth, and small grain sizes elevate the opacity away from the mid-plane. Self-consistent calculations including radiative heat transfer and condensation theory suggest that the mid-disk temperature could be high enough for silicates to remain evaporated longer than previous estimates. The formation of a condensation front leads to contrasting settling behaviors between highly refractory elements, such as Al and Ca, and moderately refractory elements, such as Mg, Si, and Fe, suggesting that elemental abundance in planetesimals may not be a simple function of volatility. 
\end{abstract}

\section{Introduction}
Astrophysical studies since the 1960s have provided a general theoretical framework for the physics of disk evolution and planetesimal formation. The core accretion model proposed by \cite{Safronov1972} and \cite{Hayashi1985}, for example, has become part of the commonly adapted theory for explaining planetary formation (e.g., \citealp{Chambers2014}). However, the problem of meter-size barrier, i.e., the infall of dust caused by headwind \citep{Weidenschilling1977}, still remains a considerable impediment in the theory of planetary formation. It is crucial to understand how planetesimals are created out of newly condensed dust particles, because the failure of forming planetesimals would leave little chance for subsequent planetary formation. A number of models have been proposed to overcome this difficulty, including efficient sticking \citep{Weidenschilling2011}, mid-plane gravitational instability \citep{Youdin2002, Chiang2008}, and streaming instability \citep{Youdin2005, Johansen2007}, but no consensus has been reached yet. Different mechanisms could result in disparate chemical structures and thus different implications for cosmochemistry \citep{Cassen1996, Ciesla2008}.

The very first prerequisite for planetesimal formation is the settling of particles towards the disk mid-plane. Most of the models proposed to overcome the meter-size barrier require a higher dust-to-gas ratio than expected from the solar abundance of elements, and dust concentration through vertical settling is one of the source mechanisms \citep{Goldreich1973, Chiang2008}. Evaluating the time scale for settling is thus essential when considering the subsequent evolution. Calculations assuming a laminar disk suggest the time scale of $\sim$10$^3$~years, which is shorter than that of radial disk evolution \citep{Safronov1972, Nakagawa1986}. When disk turbulence is included, however, dust growth is strongly inhibited, and dust grains may not settle for a long period of time \citep{Weidenschilling1984, Ciesla2007}.

The effect of chemistry on settling, however, has not been considered in previous studies, although it could play an important role in an early evolutionary stage. When dust grains settle through a vertical temperature gradient, the stable phases of dust-composing elements are likely to change, and even the existence of dust itself is not always guaranteed. Using a thermodynamic database, condensing species can be calculated at given pressure and temperature \citep{Grossman1972, Yoneda1995, Lodders2003}. This allows us to predict the compositional evolution of dust grains, but previous studies on dust settling have rarely incorporated this thermodynamic constraint.

There exist some efforts to incorporate chemical effects into the evolution of protoplanetary disks \citep{Cuzzi2004, Cassen2001, Ciesla2008, Estrada2016}, but the focus of such studies is on radial evolution, and an accurate understanding of the dynamics of vertical dust settling is difficult to be gleaned from them because of simplifying assumptions employed.  \cite{Cassen2001}, for instance, treated condensation in a 2-D dynamic model through condensation temperature to provide an explanation for the elemental abundances observed in chondrites, but the use of condensation temperature, which is constant for each element, cannot account for the dynamic nature of condensation in an evolving chemical environment. Turbulent mixing and the dependence of opacity on dust/gas ratio and grain size are likely to play an important role as well, both of which are ignored in those studies. Relevant previous studies are discussed in some detail in Section~4.3.

By building a simple yet thermodynamically-consistent model, we will quantify the effects of chemistry on vertical dust settling in protoplanetary disks, which is one of the most fundamental processes in planetary formation. Our approach is notable at least for the following two aspects. First, dust properties are calculated through Gibbs free energy minimization rather than using condensation temperature. Second, the disk thermal structure is calculated using the opacity information consistent with results from Gibbs free energy minimization. The existence of dust will modify the opacity by orders of magnitude, thus affecting the overall temperature structure. Such a temperature change will feed directly back to the chemistry, possibly altering the stable phases, and further modifying the opacity. Thus, opacity is a key parameter connecting physics and chemistry, being crucial to making our calculations fully self-consistent. The purpose of this study is to demonstrate quantitatively this feedback between physics and chemistry during the dust settling process. The paper is organized as follows. First, a theoretical formulation for the interaction between mechanics and condensation is described in detail. Modeling results then follow, exhibiting profound differences from classical calculations. Implications for astrophysical and cosmochemical studies are discussed, including the time scale for dust settling and the trends of element abundance recorded in chondrites. Previous radial evolution models involving both astrophysics and cosmochemistry are discussed as well.

\section{Method}
Our model calculates one-dimensional (1-D) dust settling, tracking the temporal and spatial evolution of dust amount and composition at an early stage of a protoplanetary disk in the terrestrial region.
Viscous dissipation is likely to cause an increase in temperature towards the mid-plane in the vertical direction. The chemical compositions of dust and coexisting gas at different heights will reflect such a temperature variation, and the amount of dust and its composition will also evolve as dust grains settle and diffuse. Our calculations are centered on the following radiative heat transfer equation \citep{Cassen2001}:
\begin{equation} \label{dT/dz}
	\frac{dT}{dz} = - \frac{3 \overline{\kappa_R} \rho_g}{16 \sigma_B T^3} F_z,
\end{equation}
where $T$ is the temperature, $z$ is the height from the mid-plane, $\overline{\kappa_R}$ is the Rosseland mean opacity, $\rho_g$ is the background gas density, $\sigma_B$ is the Stefan-Boltzmann constant, and $F_z$ is the vertical radiative flux. The Rosseland mean opacity is sensitive to dust amount, and the physics and chemistry of dust settling are connected through this opacity factor.

Our model considers a system spanning from the mid-plane to five times the initial pressure scale height. At the outer edge of our model, therefore, the initial gas density is only 10$^{-6}$ of that at the mid-plane, and the temperature is expected to become roughly constant above the outer edge. The model is spatially discretized uniformly; density, temperature, and gas and dust compositions are calculated for each grid cell. The mass and composition of dust are solved using the condensation theory. In our model, the equation of motion, the radiative heat transfer equation, dust opacity, and Gibbs free energy minimization are solved sequentially so that the distribution of dust is consistent with its thermodynamical stability. At each time step, we (1) settle and diffuse dust grains, and (2) solve radiative heat transfer and Gibbs free energy minimization simultaneously to obtain a new temperature profile and chemical composition. The detailed description of our modeling procedure is given below.

\subsection{The Motion of Dust Grains}
The motion of dust is affected by vertical settling and turbulent diffusion, as described with the following advection-diffusion equation \citep{Ciesla2010},
\begin{equation} \label{addf}
	\pd{\rho_i}{t} = - v_{sett} \pd{\rho_i}{z} + \pd{}{z} \left[ \nu \rho_g \pd{}{z} \left( \frac{\rho_i}{\rho_g} \right) \right],
\end{equation}
where $\rho_i$ is dust density of species $i$, $t$ is time, $v_{sett}$ is the vertical settling velocity, and $\nu$ is the kinematic viscosity. The dust density here refers to the total mass of dust species $i$ in a unit volume of gas, and it is different from the material density of dust grains. Equation~(\ref{addf}) has to be solved together with the background gas density distribution $\rho_g(z)$, as described in Section~2.6. The first and second terms on the right-hand side represent, respectively, advection by vertical dust settling and diffusion by turbulence. The settling of $\mu$m-size dust grains is controlled by the solar gravity and the Epstein drag \citep{Nakagawa1986}. All the dust grains are assumed to be spherical and have no porous space inside, to focus on the effect of condensation. For an early-stage, high-temperature regime under consideration, this assumption is likely to be valid. The stopping time, the time for dust particles to reach the terminal settling velocity, is inversely proportional to the background gas density. The stopping time for the grain size of $\sim$1~$\mu$m is on the order of few hours even at the low gas density region furthest away from the mid-plane. Therefore, all the grains are assumed to settle in the terminal velocity proportional to grain size as given by \citep{Goldreich1973}
\begin{equation} \label{vset}
	v_{sett} = - \frac{\rho_m}{\rho_g} \frac{s}{v_{th}} \Omega_K^2 z,
\end{equation}
where $\rho_m$ is the material dust density, $s$ is the grain size, and $\Omega_K$ is the local Keplerian angular velocity. The mean thermal speed of molecules is given by $v_{th} = \sqrt{8k_BT / \pi m_g}$, where $m_g$ is the mean gas molecular mass and $k_B$ is the Boltzmann constant.

\subsection{Turbulent Diffusion}
The turbulent diffusion in Equation~(\ref{addf}) assumes that the diffusive flux of dust grains is proportional to their concentration gradient. The kinematic viscosity of the gas is used as a diffusion coefficient, because the turbulent motions of gas and dust are well-coupled when the stopping time of dust grains is shorter than the orbital period \citep{Youdin2007}. The gas viscosity is scaled using the $\alpha$-prescription of \cite{Shakura1973} as
\begin{equation}
	\nu = \alpha c_s H = \alpha c_s^2 \frac{1}{\Omega_K},
\end{equation}
where $c_s$ is the isothermal sound speed given by $\sqrt{k_B T/m_g}$, $H$ is the pressure scale height, and $\alpha$ is the constant to account for the undetermined mechanism for viscosity. The observations of T-Tauri stars suggest $\alpha \sim 0.01$ \citep{Calvet2000}, whereas the recent studies of purely hydrodynamic turbulence imply $\alpha \sim 4 \times 10^{-4} - 10^{-3}$ \citep{Nelson2013, Stoll2014}. We employ values between $\alpha = 10^{-4}$ and $10^{-2}$ to account for this uncertainty. Whereas the value of $\alpha$ could vary vertically depending on the source mechanism of viscosity as well as the existence of dead zone \citep{Fleming2003}, a constant value of $\alpha$ along all height is adopter here. During the early stage of disk evolution, the terrestrial region is hot enough so that the gas is partially ionized, and vertically uniform turbulence is likely to be generated. In addition to dust, the composition of gas is diffused by turbulence as well. The evolution of gas is modeled also by Equation~(\ref{addf}) but with $v_{sett} = 0$.

\subsection{Temperature Structure}
\begin{figure}[ht!]
\epsscale{0.9}
\plotone{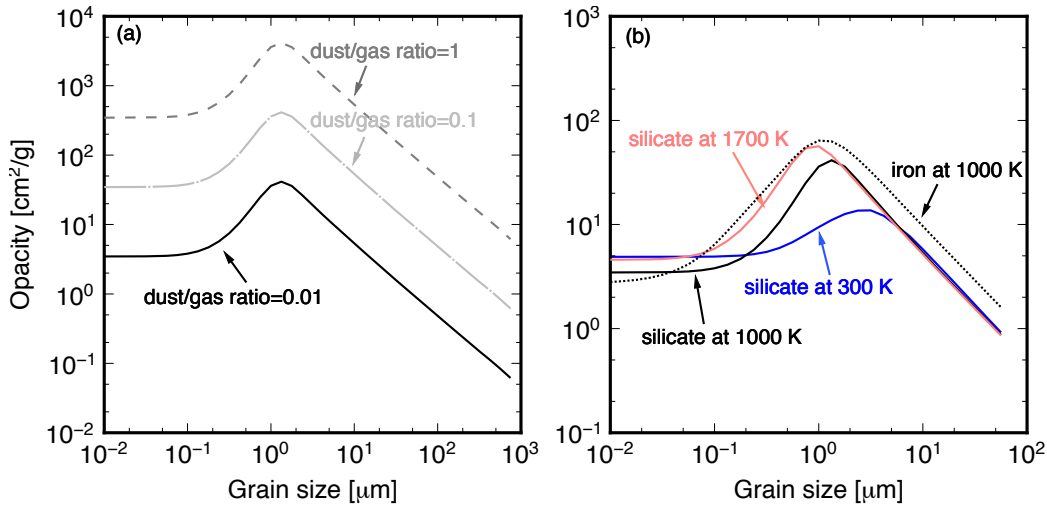}
\caption{(a) The Rosseland mean opacity of silicates at 1000~K as a function of dust particle radius. Calculations for solar composition with dust/gas mass ratio of 0.01 (solid line), 10 times enhancement of the dust/gas ratio (dotted line), and 100 times enhancement (dashed line) are shown. (b) Same as (a), but changing temperature and composition of dust grains. Silicates at 300~K (blue), 1000~K (solid black), and 1700~K (red), and iron at 1000~K (dotted black) are plotted. Dust/gas mass ratio of 0.01 is used here.}
\label{fig_op}
\end{figure}

The temperature distribution is solved using the radiative transfer equation: Equation~(\ref{dT/dz}) supplemented with
\begin{equation} \label{dFz/dz}
	\frac{dF_z}{dz} = \frac{9}{4} \nu \rho_g \Omega_K^2.
\end{equation}
The Rosseland mean opacity of dust particles is calculated as a function of the dust/gas ratio, grain size, and temperature through the Mie-scattering code of \cite{Matzler2002} using the refractive index data of silicates \citep{Draine1985}. The difference of refractive indices between various dust species is not treated here. The opacity varies primarily with the amount of dust and its grain size \citep{Pollack1994}. The Rosseland mean opacity is plotted as a function of grain size in Figure~\ref{fig_op}(a). Dust particles of 1~$\mu$m size have the highest opacity for near-infrared radiation. The opacity is also affected by temperature and dust composition, though their influence is minor compared to that of dust/gas ratio as shown in Figure~\ref{fig_op}(b).

The grain size affects both opacity and settling speed, playing a crucial role in our model. Dust grains are assumed to nucleate homogeneously when the dust first condenses. \elm{and to grow heterogeneously when existing grains are present. } For the initial grain size, we test a range of values spanning from 0.1~$\mu$m to 10~$\mu$m. The lower bound is taken from a typical interstellar grain size, and the upper bound is estimated quantitatively from the mean free path, $\lambda$, and the thermal velocity of molecules $v_{th}$ assuming perfect aggregation at each collision. By perfect aggregation, we mean that two molecules or particles collide at an interval of $\lambda/v_{th}$, and they always form an aggregate without bouncing. As molecules collide and stick, the number density becomes lower and the interval of collision longer. One-hundredth of the time step (see Section~\ref{imple}) is considered as the nucleation time scale, which is $\sim$5 to 100 years in our calculations, and this gives the upper bound of 10~$\mu$m as an initial nucleation size. Given the assumption of perfect aggregation, it should be considered as an unlikely end-member case. This range mostly covers the typical grain size of matrix materials (50~nm - 5~$\mu$m) in chondrites \citep{Alexander1989, Scott2014}, which could be considered as the typical size of unmelted grains. \edy{Grain growth in our model occurs through additional condensation on existing grains, i.e., the number of dust particle is conserved by assuming heterogeneous nucleation. However, it only changes particle radius by a factor of two at most, and therefore our results are characterized mostly by the choice of initial nucleation size.} Grain fragmentation, sticking, and dust size distribution are not considered in our model, and this simplification is likely to be justified a posteriori for most cases, as discussed in Section~3.

Equations (\ref{dT/dz}) and (\ref{dFz/dz}) are integrated from the disk surface towards the mid-plane. The surface temperature, $T_s$, is used as a boundary condition and is converted from the effective temperature of the disk, $T_{e}$, assuming that the disk is opaque to blackbody radiation due to the existence of $\mu$m-size grains. By balancing viscous dissipation and blackbody radiation, these temperatures are given by
\begin{equation} \label{teff}
	T_s = \frac{T_{e}}{2^{1/4}}
\end{equation}
and
\begin{equation} \label{balance}
	\sigma_B T_e ^4 = \frac{9}{8} \nu \Sigma \Omega_K^2,
\end{equation}
where $\Sigma$ is the disk surface density \citep{Cassen1996, Chiang1997}. The viscosity at the mid-plane is used for $\nu$ in Equation~(\ref{balance}).

\subsection{Condensation}
The amounts of dust and its average chemical compositions at different heights in the system are calculated through Gibbs free energy minimization. The disk temperature and pressure are necessary to determine stable phases, and the pressure is calculated assuming the ideal gas law ($P = \rho_g k_B T/m_g$). Previous settling models have fixed the total dust amount in the system (e.g., \citealp{Nakagawa1986, Ciesla2007}), but our model determines the amount based on condensation calculation, making the model consistent with evolving chemistry. To facilitate our modeling effort, we have developed an original optimization method to minimize the free energy using a non-linear conjugate gradient method (see Appendix A for detail). Only major elements (H, O, Na, Mg, Al, Si, Ca, and Fe) are considered in equilibrium calculations in this study. Mineral species considered include corundum, melilite, spinel, olivine, and pyroxene. The complete list of species considered is given in Table~\ref{spec}. Thermodynamical data for silicates phases are taken from \cite{Robie1995}, and those for the rest of phases are from the JANAF Thermochemical Tables.

\subsection{Implementation} \label{imple}
The distributions of temperature and dust composition are calculated based on the advected and diffused composition profile, using radiative heat transfer and Gibbs free energy minimization simultaneously.
At the beginning of each time step, all grids store dust compositions from the previous time step.
Gibbs free energy minimization is then performed using the temperature and pressure at the upper node of the cell in order to obtain the new dust composition consistent with the new temperature profile.
As we integrate the heat transfer equation from the disk surface to the mid-plane, the upper node is always updated first, and the upper node temperature corresponds to the new temperature for the cell. The temperature of the lower node of that cell is computed by integrating Equations~(\ref{dT/dz}) and (\ref{dFz/dz}), using the opacity based on the dust amount obtained from free energy minimization.

When a grid cell has high dust density, a high opacity value of the cell could result in a substantial temperature difference between the upper and lower nodes of the grid. Dust grains may not be able to remain in a solid phase anymore and may evaporate at the lower end of the cell. However, using a single high opacity value does not account for this compositional variation and would result in an unrealistically high temperature at the lower node. In this case, we subdivide the grid cell and recalculate temperature and composition with higher resolution. We continue this grid refinement until the additional subdivision changes the temperature of the lower node by less than 1~K. In this manner, the dust profile will be consistent with both radiative heat transfer and chemical thermodynamics. We used 100~grid cells (size of 0.05~scale height), and this particular choice of discretization is sufficient for our purpose, as shown later. The time step, $\Delta t$, is chosen so that $\Delta t < 0.1 \Delta z/ v_{sett}$ for the bottom 95\% of dust grains by mass.

\subsection{Background Gas Density Evolution}
The surface density $\Sigma$ is solved using a 1-D radial disk evolution model, derived from mass and angular momentum conservation laws,
\begin{equation} \label{sigma}
	\pd{\Sigma}{t} = \frac{3}{r} \pd{}{r} \left[ \sqrt{r} \pd{}{r} (\nu \Sigma \sqrt{r}) \right],
\end{equation}
where $r$ is the radial distance from the Sun \citep{Pringle1981}. When solving this equation, the viscosity~$\nu$ at the mid-plane is used, which is again described by the $\alpha$-prescription. When using Equation~(\ref{sigma}), small differences in mid-disk temperature do not lead to an appreciable change in gross radial evolution, so we compute the mid-disk temperature, $T_{mid}$, in an approximated way as
\begin{equation} \label{Pollack}
	T_{mid}^4 = \frac{3}{2} \tau_0 T_s^4,
\end{equation}
where $\tau_0$ is the optical depth between the surface and the mid-plane. The optical depth is calculated as $\overline{\kappa_R} \Sigma/2$, \edy{where the Rosseland mean opacity at the mid-plane is calculated to be consistent with the average dust/gas ratio of the vertical column \citep{Ruden1991, Cassen1996}}. As our model results with thermodynamic calculations show later (Section~4.1), the mid-disk temperature could be higher by up to 400~K than predicted by Equation~(\ref{Pollack}).

The vertical gas density profile is assumed to satisfy a hydrostatic equilibrium, and the surface density is converted to the vertical profile through
\begin{equation} \label{hydro}
	\rho_g(z) = \rho_0 \exp \left(- \frac{z^2}{2 H^2} \right),
\end{equation}
where $\rho_0$ is the density at the mid-plane defined as $\Sigma/(\sqrt{2\pi}H)$. The background gas density is assumed to vary temporally in proportion to the change of the surface density by infall to the Sun or by photoevaporation \citep{Shu1993}. In our model calculations, the gas composition is modified only through vertical processes such as condensation, vertical dust settling, and turbulent diffusion, and we do not explicitly model the effect of radial transport on the gas composition. This is equivalent to assuming a similar chemical composition in the nearby region, and thus the radial mass flux into the system would not modify the gas composition. Assuming a hydrostatic balance, Equation~(\ref{addf}) may be expressed as
\begin{equation} \label{addf_imp}
 	\pd{\rho_i}{t} = - \left( v_{sett} + v_{hydro} \right)  \pd{\rho_i}{z} +  \pd{}{z} \left( \nu \pd{\rho_i}{z} \right)
\end{equation}
where $v_{hydro} = - (\nu z)/H^2$ is the effective velocity that suppresses the dust motion to diffuse away from the mid-plane, originated from the decrease in the gas density. This equation is used to track the evolution of dust density in our implementation.

\section{Results}
\begin{figure}[ht!]
\epsscale{1}
\plotone{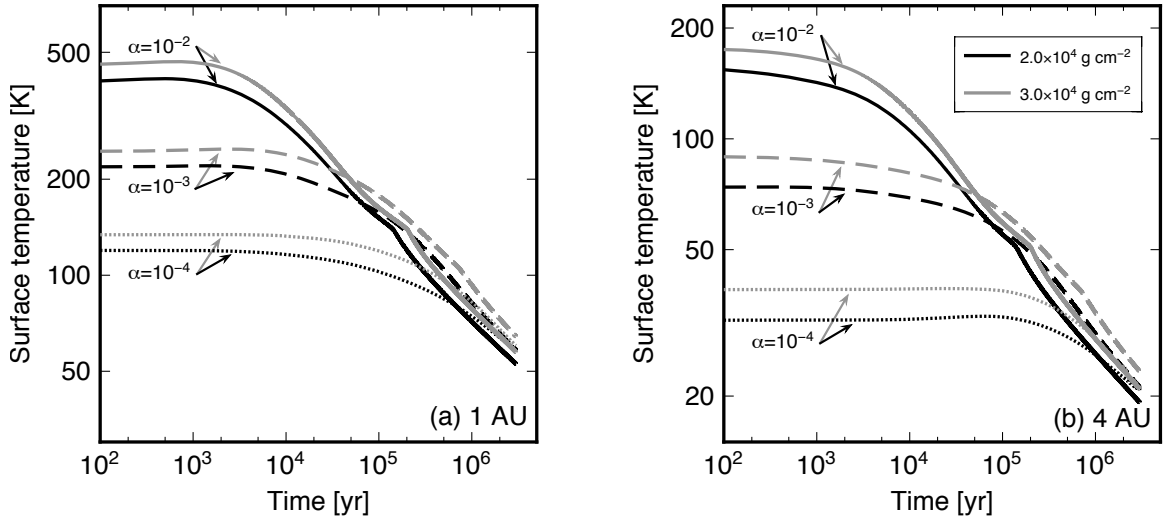}
\caption{The evolution of surface temperature at (a) 1~AU and (b) 4~AU. Six model runs with different values of $\alpha$ and surface density are plotted. The values of $\alpha$ and $\Sigma$ are denoted by line types and colors, respectively, as $\alpha=10^{-2}$ (solid), $10^{-3}$ (dashed), and $10^{-4}$ (dotted), and $\Sigma_0 = 2.0 \times 10^4$~g~cm$^{-2}$ (black) and $3.0 \times 10^4$~g~cm$^{-2}$ (grey). The time scale of evolution changes linearly with the value of $\alpha$.}
\label{fig_tE1}
\end{figure}

The evolution of gas surface density is calculated first using the radial disk evolution model of Equation~(\ref{sigma}), before solving the settling. The surface temperature is converted from the surface density using Equations~(\ref{teff}) and (\ref{balance}), and its evolution at 1~AU and 4~AU is plotted in Figure~\ref{fig_tE1}. The time scale for disk evolution decreases inversely with the value of $\alpha$, which is proportional to viscosity. The evolution also depends strongly on an initial density profile, the exact form of which has long been debated. In this study, we test two initial disk masses, 0.21 and 0.32~solar mass, with a surface density profile inversely proportional to the distance from the Sun \citep{Hartmann1998}, spanning up to 15~AU ($\Sigma(t=0,r) = \Sigma_0/r$, where $\Sigma_0$ is the surface density at 1~AU). We used three values of $\alpha$ ($10^{-4}$, $10^{-3}$, and $10^{-2}$) combined with two values of surface density $\Sigma_0$ (2.0 $\times$ 10$^4$ and 3.0 $\times$ 10$^4$~g~cm$^{-2}$, corresponding to 0.21$M_\odot$ and 0.32$M_\odot$, respectively). 

\begin{figure}[ht!]
\figurenum{3}
\epsscale{0.9}
\plotone{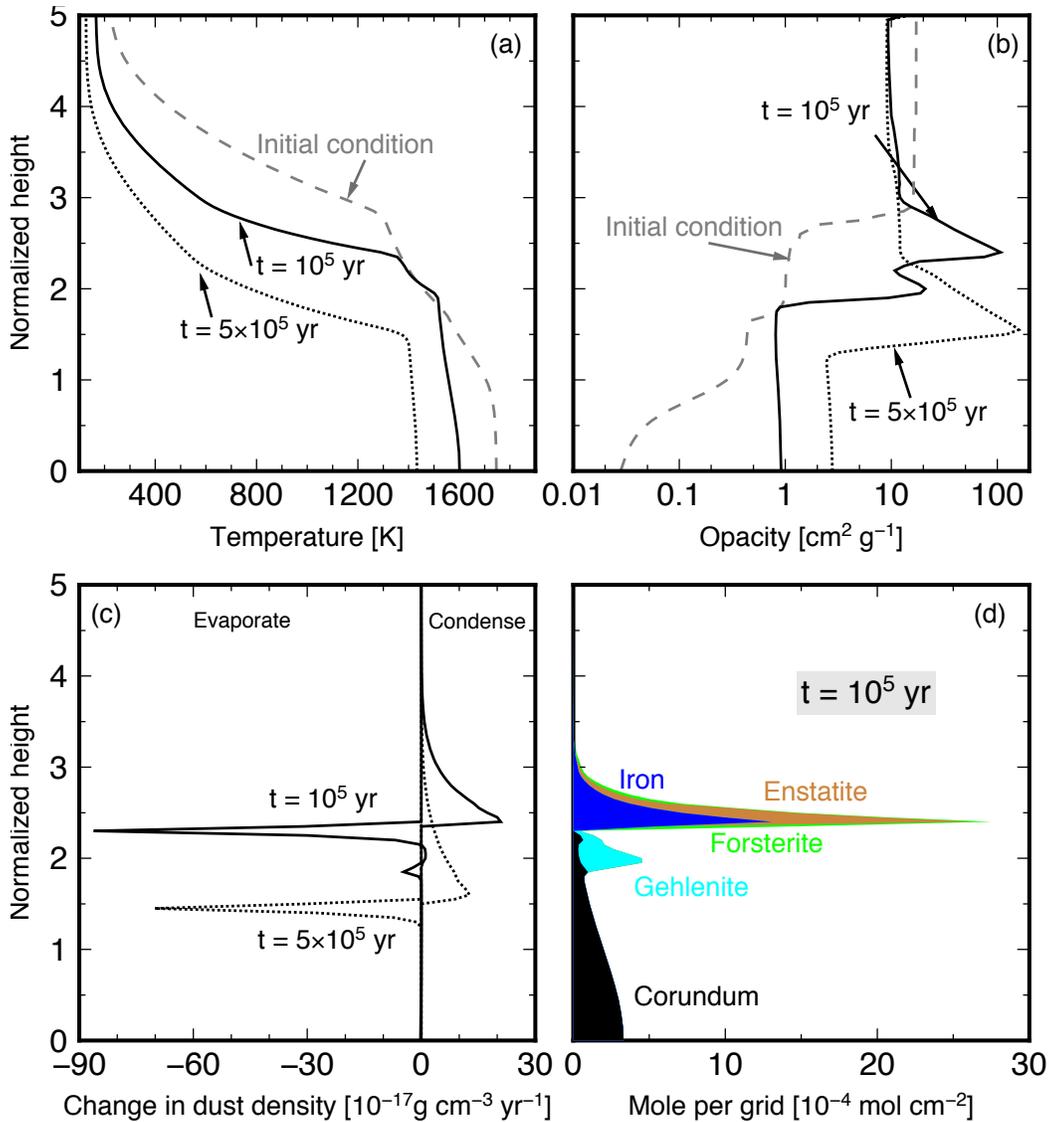}
\caption{Model results at 1~AU with $\alpha=10^{-3}, \Sigma_0=2\times10^4$~g~cm$^{-2}$, and initial nucleation size of 1~$\mu$m. (a) Temperature profiles at $t = 0$ (dashed grey), $10^5$ (solid black), and $5 \times 10^5$ years (dotted black). (b) Corresponding profiles of opacity. (c) Change in dust density due to condensation or evaporation at $t = 10^5$ (solid) and $5 \times 10^5$~years (dotted). A pair of evaporation and condensation peaks creates a condensation front. (d) Aggregated mass distribution at $t = 10^5$ yr. Individual distributions of different dust species are denoted by different colors. The opacity peak in (b) at $\sim$2.3 scale height correspond to the concentration of metallic iron, forsterite, and enstatite. Akermanite, spinel, fayalite, ferrosilite, and feldspar also exist in smaller quantities, but they are not shown for the sake of clarity. \edy{The temperature inflection in (a) indicates the condensation temperature of silicates.}}
\label{fig_set1}
\end{figure}

We calculated the behavior of dust settling at 1~AU to 5~AU with various parameter combinations. Our model starts with the solar composition of \cite{Lodders2003} at all heights. When the disk interior is hot enough, Si, Mg, and Fe evaporate, and when it is even hotter, Al and Ca also start to exist as gas phases.
As we are considering an early stage of disk evolution, viscous heating is a major heat source of the disk, and the radiation from the Sun has a negligible effect. Depending on the value of $\alpha$, viscous heating is one to three orders of magnitude larger than the heat flux from the solar radiation, which is 3$\times 10^3$~erg~cm$^{-2}$~s$^{-1}$ at 1~AU assuming a flat disk \citep{Chiang1997}. This leads to an increase in temperature towards the mid-plane. Dust first precipitates in the cooler upper region of the disk, and it sinks toward the mid-plane due to the gravity from the Sun. The settling of $\mu$m-size particles is slow, but it gradually increases the dust/gas ratio in the lower region.
In the inner region of the disk, if the value of $\alpha$ is high enough, dust grains evaporate as they descend into the hotter interior and are prevented from further settling and grain growth. Figure~\ref{fig_set1} shows some snapshots for the profiles of temperature and dust composition.

The evaporation of dust particles creates a concentration of dust-composing elements. This leads to the formation of a dust-rich layer with high opacity, creating a large temperature gradient. Any dust grain trying to cross this temperature will be subjected to evaporation, which prevents the grain from settling further (Figure~\ref{fig_set1}(a)). This forms a ``condensation front'', where dust particles concentrate due to a temperature increase created by themselves. Figure~\ref{fig_set1}(d) shows that Mg, Si, and Fe are concentrated above the condensation front at $\sim$2.3~times the scale height at $t = 10^5$ year, because these elements all condense in a narrow temperature range. Figure~\ref{fig_set1}(c) shows a pair of condensation and evaporation rate peaks occurring above and below the front. The condensation peak is stretched above the front because of the turbulent diffusion. For the purpose of quantification, a condensation front is defined to have formed when total mass of evaporated silicates is larger than 10~\% of that of newly condensed silicates. All of our calculations were made with 100~grid cells, but a finer grid resolution does not change our results significantly. Table~\ref{conv} compares the values of the mid-disk temperature at $t=5\times 10^4$ years obtained by varying the grid size.

\begin{figure}[ht!]
\epsscale{0.9}
\figurenum{4}
\plotone{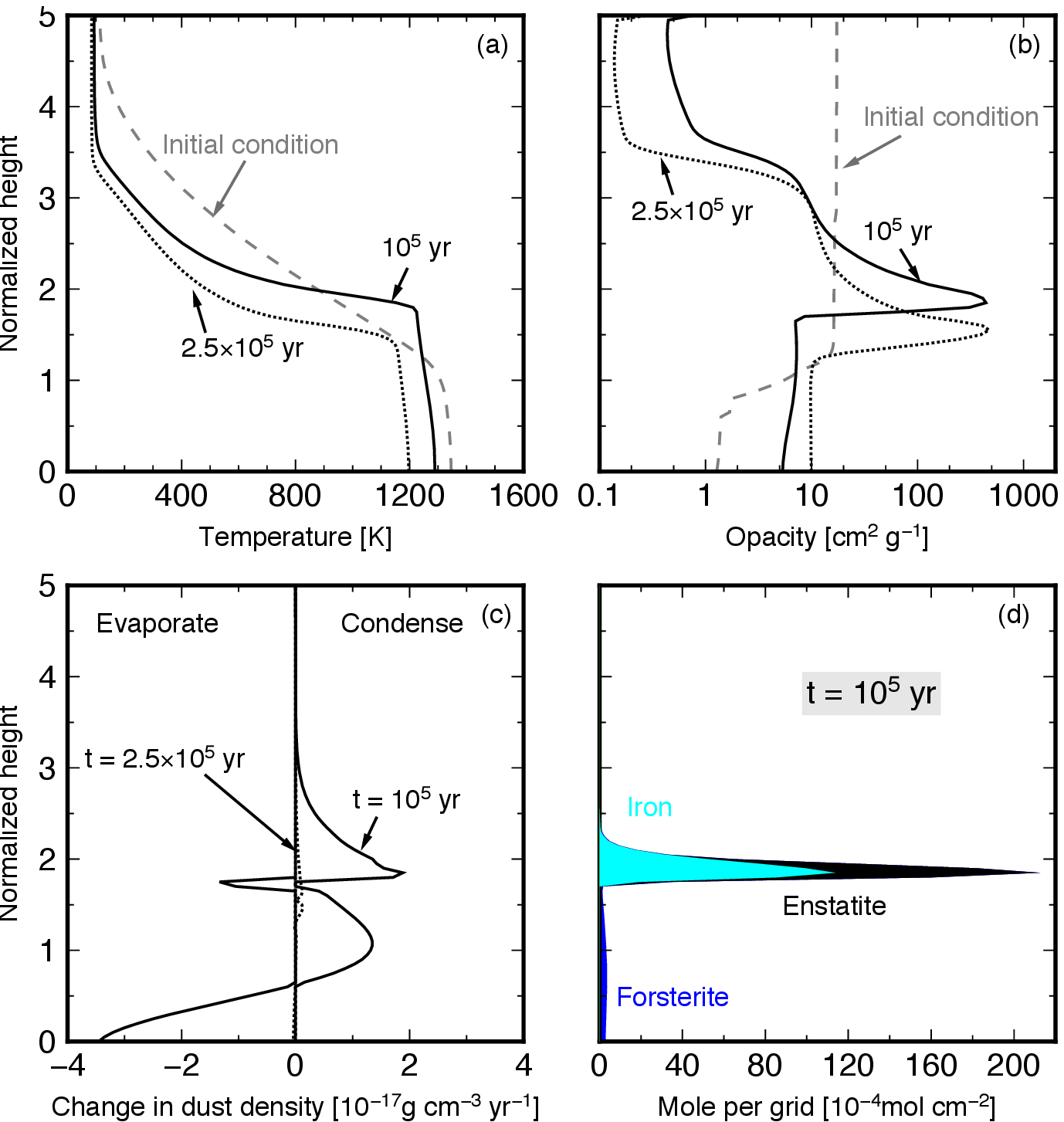}
\caption{Same as Figure~\ref{fig_set1}, but at 3~AU for $t = 0$ (dashed grey), $10^5$ (black), and 2.5$\times10^5$~years (dotted black). At $t=10^5$~years, forsterite has started to condense near the mid-plane, which indicates that the disappearance of condensation front in a short time. Indeed, by $t = 2.5 \times 10^5$~years, the condensation front has disappeared almost entirely.}
\label{fig_set3}
\end{figure}

Disk turbulence would transfer the gas and dust grains away the front both upwards and downwards, and dust-composing elements, Mg, Si, and Fe, will experience repeated evaporation and condensation every time dust crosses the front. This keeps the grain size close to its initial nucleation size. Some grains might go through collisional sticking and grow in size, but such larger grains would settle faster into a hotter region, ending in evaporation. The condensation front and the associated temperature jump will be sustained by the remaining smaller particles, justifying our strategy of neglecting grain growth in our calculations. The opacity model including particle sticking and fragmentation could play a key role in the environment where evaporation and condensation are not happening, and in such environment, the grain size is likely to follow a power-law distribution \citep{Williams1994, Birnstiel2011}. Near a condensation front, however, evaporation and condensation are likely to happen so frequently that grain size is expected to be determined mainly by the speed of nucleation. In other words, the grain size distribution will have a characteristic size determined by nucleation. The small grain size, through its influence on opacity, helps to maintain high mid-disk temperature (Figure~\ref{fig_set1}(a) and \ref{fig_set3}(a)). \edy{Some studies (e.g., \citealp{Weidenschilling1993}) suggest that the aggregation of $\mu$m-size particles could occur rapidly on the order of $10^3 - 10^5$~years. Such aggregates would have low opacity, yet their settling is slow due to its porous structure \citep{Ormel2007}. These grains, however, are likely to evaporate quickly because turbulent mixing in the vertical column occurs in a much shorter time scale of $H^2/\nu$ ($\sim10$~years).}

Both the surface and mid-disk temperatures gradually decrease as viscous heating diminishes along with the dissipation of the nebular gas. The disk interior, where most of the mass exists, remains hotter than the condensation temperatures of Mg, Si, and Fe due to the condensation front, whereas the region closer to the surface cools down within a few thousands of years. In all cases, the mid-disk temperature does not become high enough to evaporate highly refractory species (corundum and melilite). The amount of Al- and Ca-bearing grains is not high enough to affect the thermal structure; they only condense in the hot inner region, where it has too little radiative heat flux to create a large temperature gradient. Highly refractory species are likely to grow in size without evaporating, and their evolution can be treated using a traditional model. The grain-size evolution should be modeled in a more careful manner for these species \citep{Estrada2016}.

\begin{figure}[tbh!]
\epsscale{1.2}
\figurenum{5}
\plotone{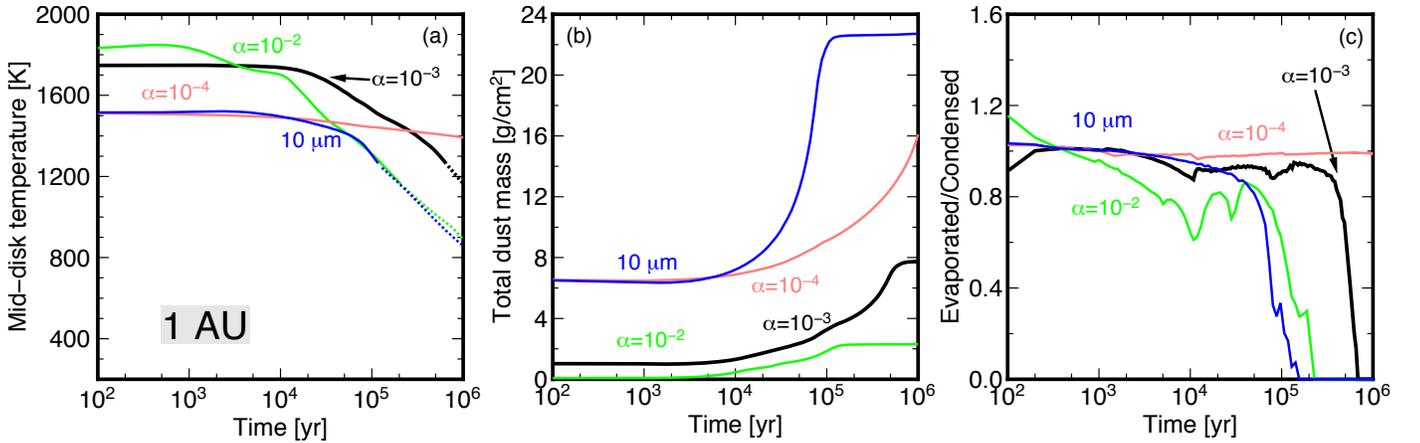}
\caption{Evolution at 1~AU of (a) mid-disk temperature, (b) total dust mass, and (c) the ratio of newly evaporated and condensed amounts in a given time step, for the model with $\alpha=10^{-3}$, $\Sigma_0=2\times10^4$~g~cm$^{-2}$, and the initial nucleation size of 1~$\mu$m (solid). Effects of modifying one of those parameters are also shown: $\alpha=10^{-2}$ (green), $\alpha=10^{-4}$ (red), and the initial nucleation size of 10~$\mu$m (blue). A larger viscosity causes more dust diffusion, and a larger initial nucleation size would causes a colder interior, but they still exhibit the feature of a condensation front, indicating that the condensation front is ubiquitous at 1~AU. \edy{In (a), the thermal evolution after the disappearance of condensation front is shown with dotted lines to indicate that our modeling results are to be regarded as an upper bound on a more likely thermal evolution. Opacity will be reduced by grain aggregation when silicates become stable at the mid-plane, but such grain growth is not included in our model.}}
\label{fig_tev}
\end{figure}

\begin{figure}[tbh!]
\epsscale{1.2}
\figurenum{6}
\plotone{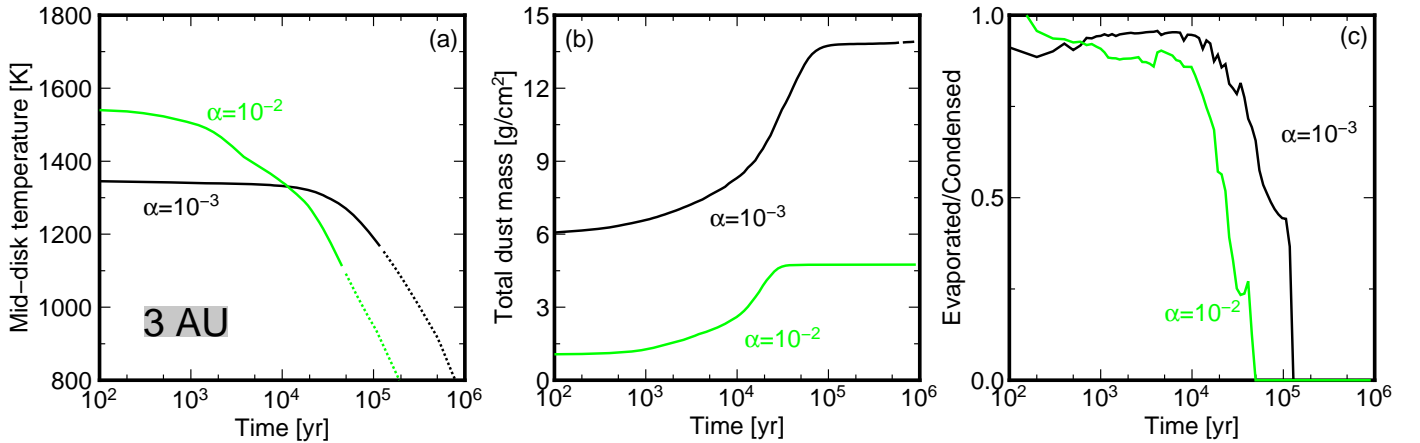}
\caption{Same as Figure~\ref{fig_tev} but at 3~AU. $\alpha=10^{-3}$ (black) and $10^{-2}$ (green) with $\Sigma_0=2\times10^4$~g~cm$^{-2}$ and the initial nucleation size of 1~$\mu$m. For $\alpha=10^{-2}$, the mid-disk temperature starts to decrease as the surface temperature decreases with the radial evolution (Figure~\ref{fig_tE1}), whereas in the case of $\alpha=10^{-3}$, it remains high for a more prolonged period.}
\label{fig_tev3}
\end{figure}

Representative cases of temporal evolution at 1~AU are shown in Figure~\ref{fig_tev}. As time proceeds, the average temperature decreases, and dust mass increases. The amount of dust is self-regulated below a certain value such that it does not create too large a temperature increase that prohibits the existence of the dust altogether, and the remaining dust-composing elements would remain in the gas phase. As long as the condensation front exists, \edy{the ratio of newly evaporated and condensed amount in a given time period} maintains the value close to unity (Figure~\ref{fig_tev}(c))\edy{; this approximate balance between evaporation and condensation is what maintains the condensation front at its quasi-steady state.}
Lower mid-disk temperature is observed when the initial nucleation size deviates from 1~$\mu$m or lower viscosity are adopted; the former reduces the opacity, and the latter lowers the radiative heat flux. With lower opacity or radiative heat flux, more dust is necessary to produce the high mid-disk temperature. This results in more dust to remain in the system without being lost by photoevaporation (Figure~\ref{fig_tev}(b)). When viscosity is higher, however, the mid-disk temperature decreases rapidly with time as seen in the case with $\alpha = 10^{-2}$, because the mass in the system depletes quickly due to high mass accretion rate. This is observed in all the regions from 1-3~AU (Figures~\ref{fig_tev} and \ref{fig_tev3}). The condensation front disappears when Mg, Si, and Fe-bearing species start to condense near the mid-plane (Figures~\ref{fig_set3} and \ref{fig_tev}(b)). The region near the mid-plane has lower radiative flux, resulting in a smaller thermal gradient. Therefore, a condensation front, requiring a large temperature increase, would not be able to form close to the mid-plane. Once the front disappears, dust starts to form near the mid-plane within a short period of time. 

\begin{figure}[tbh!]
\epsscale{1.2}
\figurenum{7}
\plotone{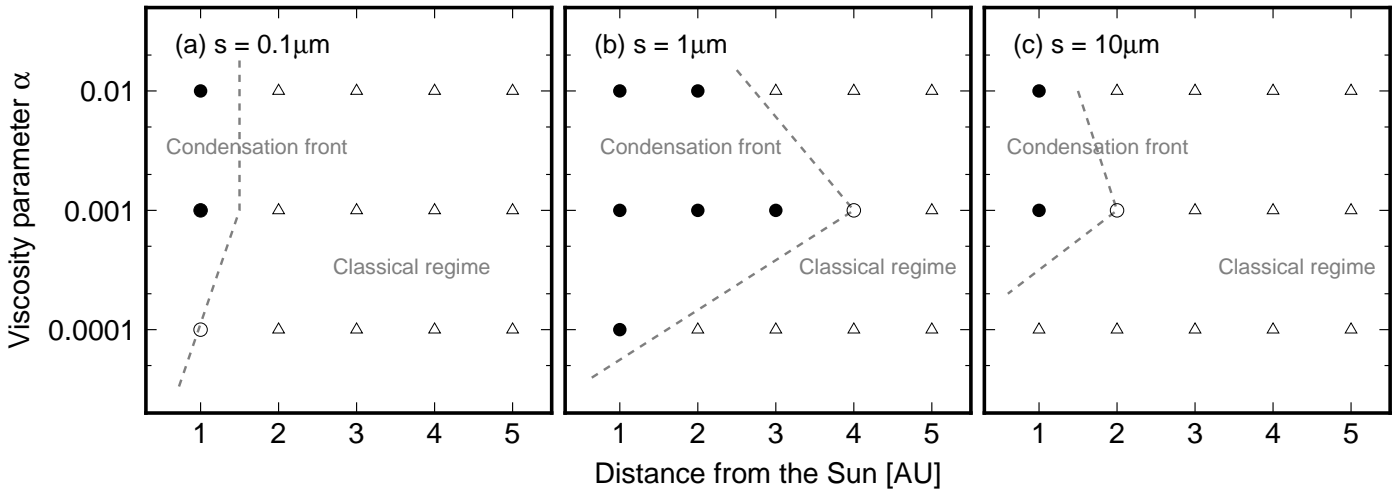}
\caption{Conditions for the condensation front regime, with the initial nucleation size of (a) 0.1~$\mu$m, (b) 1~$\mu$m, and (c) 10~$\mu$m. The classical regime refers simply to the cases in which no condensation front forms. Solid circle denotes the case of a condensation front forming under disk mass of both 0.21~$M_\odot$ and 0.32~$M_\odot$, open circle denotes that a front only forms when disk mass of 0.32~$M_\odot$, and triangle denotes that a front is not observed or \edy{it only exists for less than $10^4$ years.}}
\label{fig_sum}
\end{figure}

The formation of a condensation front inhibits dust settling and maintains a hot interior.  \edy{At 1~AU, a condensation front would last as long as $5\times10^5$~years for $\alpha = 10^{-3}$, and $10^6$~years for $\alpha = 10^{-4}$ with the initial nucleation size of 1~$\mu$m.} When a condensation front forms, a high mid-disk temperature is maintained, and silicate dust does not condense near the mid-plane until the gas phase, which is the primary source of viscous heating, dissipates along with the radial disk evolution. Whether or not dust emerges near the mid-plane has a large impact on subsequent processes (e.g., \citealp{Johansen2006}). Figure~\ref{fig_sum} summarizes the condition for the formation of a condensation front as a function of $\alpha$ and the distance from the Sun. When $\alpha$ is closer to 10$^{-3}$, a condensation front is likely to form in most of the terrestrial region. If the initial nucleation size is limited to 1~$\mu$m, the condensation front could form as far as in the asteroid belt region (Figure~\ref{fig_sum}(b)). For $\alpha = 10^{-3}$, the region spans even further close to the Jovian orbit. Even if the initial nucleation size is as small as interstellar grains or as large as 10~$\mu$m, the existence of front is likely at $\alpha \sim 10^{-3}$. The likelihood of the front formation is highest when the nucleation grain size is around 1~$\mu$m, because of the grain-size dependency of opacity (Figure~\ref{fig_op}). The actual grain size would exhibit some finite distribution, \edy{and grains with a larger fraction of mass dominate opacity in general. When grains have a distribution between 0.1 and 1~$\mu$m, opacity becomes closer to that of 1~$\mu$m. The case with the initial size of 1~$\mu$m should be regarded as a reasonable upper bound on the extent of condensation front formation. }

\section{Discussion}
\subsection{Implications for the dynamic structure in vertical direction}
Dust settling and turbulent diffusion are often the only processes considered in the evolution in vertical direction, but in the early stage of the disk evolution, the stability of dust species plays a key role in determining the structure. The existence of dust grains, which is necessary for the subsequent evolution to take place, is controlled by thermodynamics and should not be assumed a priori. In fact, in early protoplanetary disks, silicates are usually considered to be evaporated in the terrestrial planet region \citep{Bell1997}. Planetesimals start to form when silicates become stable near the mid-plane, but our modeling results suggest that the chemistry of a protoplanetary disk could have already evolved rather substantially by then if a condensation front formed. Planetesimals could have different compositions depending on which species initially exist as solid near the mid-disk.

\begin{figure}[tbh!]
\epsscale{1}
\figurenum{8}
\plotone{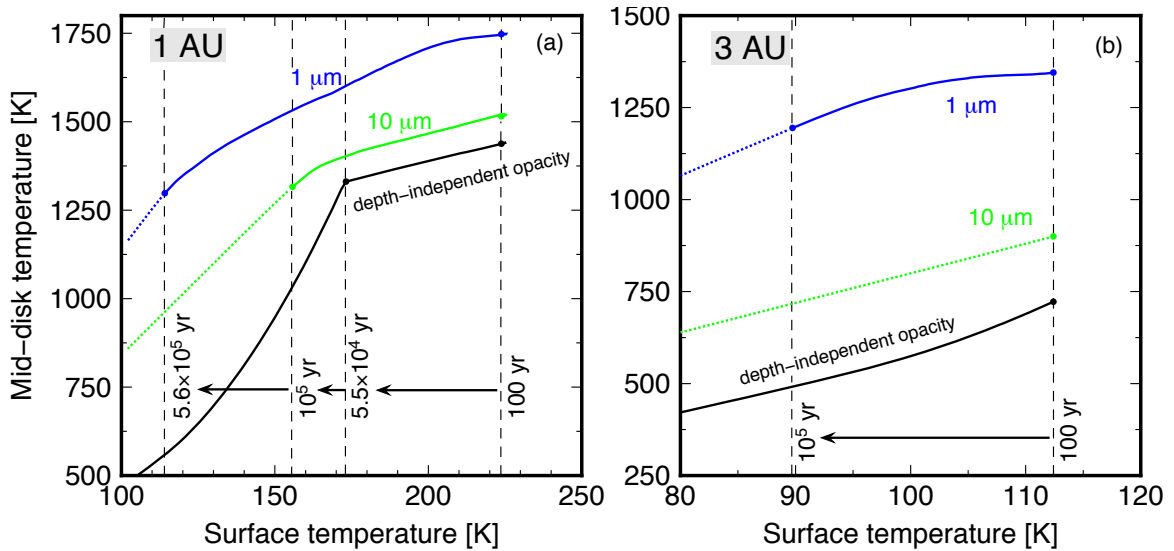}
\caption{The relation between surface and mid-disk temperatures at \edy{(a) 1~AU and (b) 3~AU with $\alpha=10^{-3}$ and $\Sigma_0=2\times10^4$~g~cm$^{-2}$. Cases with varying opacity assuming initial nucleation size of 1~$\mu$m (blue) and 10~$\mu$m (green) are compared with the case with depth-independent opacity (black). The evolution after the disappearance of condensation front is shown with dotted lines to indicate that our modeling results are to be regarded as an upper bound. For the case with depth-independent opacity, the opacity at the mid-plane is used at all height (Section~2.6). Results by \cite{Cassen1996} and \cite{Estrada2016} correspond to the case of depth-independent opacity, although their assumption of constant radiative flux predicts higher mid-disk temperature than the calculation here. The opacity of silicates is assumed to be 3.0~cm$^2$~g$^{-1}$, which is the same with our model result with 10~$\mu$m. The kink in the depth-independent case in (a) reflects the condensation of silicates. The temporal evolution of surface temperature (Figure~2) is indicated by vertical dashed lines.}}
\label{fig_surf}
\end{figure}

When condensation is considered, dust concentrates above the front forming a temperature jump, and dust would not be able to penetrate this temperature jump created by itself. Evaporation and condensation would repeat at the front, keeping the dust size in its initial nucleation size. This leads to high opacity, which maintains the mid-disk temperature hotter compared to previous calculations (\edy{Figure~\ref{fig_surf}}; \citealp{Cassen1996, Estrada2016}). A condensation front survives until gas dissipates and viscous dissipation becomes weak. Eventually, metallic iron, forsterite, and enstatite become stable near the cold mid-plane, and no further evaporation will occur. \edy{Our model assumption of high opacity due to small grain size will not be valid after this, because grains are likely to grow in size.}

In previous studies \citep{Ruden1991, Cassen1996, Ciesla2006}, the surface temperature $T_s$ and the mid-disk temperature $T_{mid}$ are related using Equation~(\ref{Pollack}). This relation assumes depth-independent opacity and radiative heat flux, but this is unlikely in the early stage of protoplanetary disk evolution.
Whereas Equation~(\ref{Pollack}) is sufficient for modeling the gross behavior of radial evolution as previously noted, \edy{it is inadequate to accurately estimate the mid-disk temperature, which is crucial to evaluate when silicates start to condense. When silicates evaporate at the mid-plane, the opacity in the interior may be low, but grains near the surface are likely to maintain small sizes, creating high opacity. Therefore, Equation~(\ref{Pollack}) underestimates the optical depth and the mid-disk temperature, because it does not consider the heterogeneous distribution of dust grains. Note that the assumption of constant vertical radiative heat flux tends to overestimate the mid-disk temperature, because the radiative flux decreases toward the mid-plane in a disk driven largely by viscous dissipation. The use of depth-independent opacity, however, underestimates the temperature to a greater extent, so these two effects do not cancel.}

\edy{The importance of vertically varying opacity may be understood from Figure~\ref{fig_surf}.} Our results indicate that the mid-disk temperature remains high enough so that silicates are evaporated for a longer time than commonly thought.  \edy{The temperature increase between the condensation front and the mid-plane cannot be described in a model with depth-independent opacity. More refractory condensates can act as an additional source of opacity to increase the mid-disk temperature when a condensation front forms further from the mid-plane. Previous models have underestimated the mid-disk temperature by neglecting the reduction of grain size through turbulence and evaporation, and such models do not accurately describe the mid-disk temperature.}

\edy{Given the temporal evolution of surface temperature (Figure~\ref{fig_tE1}), this means that dust would not settle and reach the mid-plane even after 10$^5$~years at 1~AU, which is longer than previous estimates \citep{Cassen2001, Estrada2016}, and with an initial size of 1~$\mu$m, settling may take longer than a million years (Figure~\ref{fig_tev}(c)).} In the meantime, some fraction of dust composing elements would dissipate as gas, without becoming planetesimals. This could potentially change the estimate of the minimum solar nebular mass. Incidentally, the formation ages of Ca-Al-rich inclusions and chondrites are known to be different by $\sim$1~Myr. This age gap has traditionally been explained as the melting of chondritic materials by radiogenic heat during the first one million years (e.g., \citealp{Kruijer2014}), but this could also reflect the long-lasting nature of the region where silicates evaporate \edy{at the mid-plane, because planetesimals with chondritic composition} are unlikely to form until silicates become stable near the mid-plane.

The extent of viscous turbulence could vary depending on the source mechanism, although it is simply characterized by a constant value of $\alpha$ in this study. This assumption is reasonable for most of our cases, because our focus is on the hot inner part of the disks, which remain hotter than 1000~K, leading the gas to be \edy{partially} ionized through collision. This assumption, however, would be invalid in the outer cold region. Future studies are warranted to address these important complications.

\subsection{Implications for Cosmochemical Observations}
\begin{figure}[ht!]
\epsscale{0.6}
\figurenum{9}
\plotone{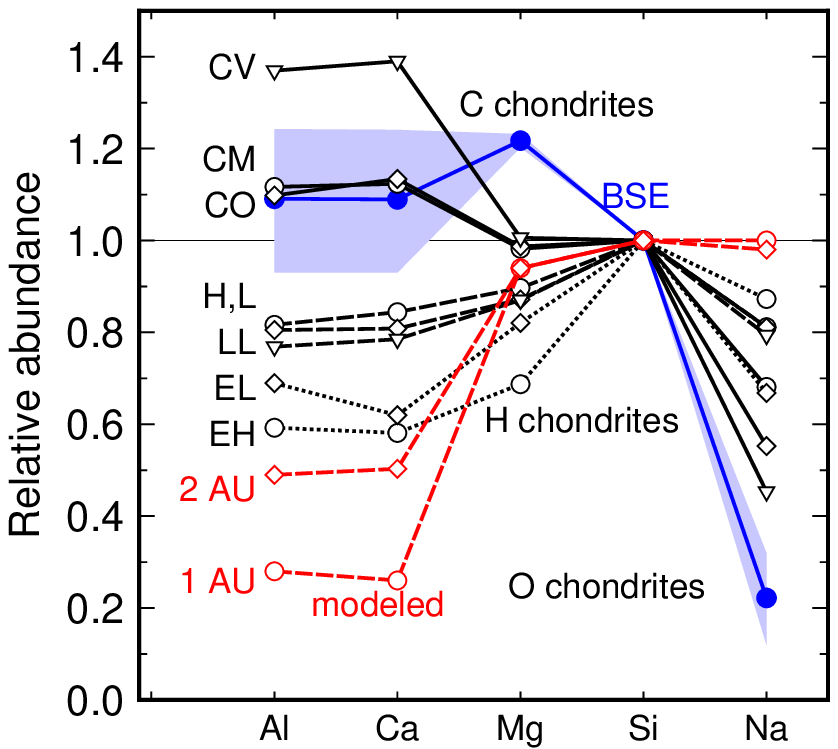}
\caption{Relative abundances of Al, Ca, Mg, and Na, normalized by Si and by CI chondrite. The abundance of BSE (blue), carbonaceous (solid), ordinary (dashed), and enstatite chondrite (dotted) are shown. The blue shaded area shows one standard deviation of the ratios. Chondrite subgroups are denoted by labels. Data for BSE and chondrites are taken from \cite{Lyubetskaya2007} and \cite{Wasson1988}, respectively. Dashed red lines show the modeled composition of predicted planetesimals at 1~AU~(circle) and 2~AU (diamond) with $\alpha = 10^{-3}$ and initial nucleation size of 1~$\mu$m. When dust grains below the front are removed towards the Sun, the resulting solid materials show a significant depletion in Al and Ca. See the text for details.}
\label{fig_me}
\end{figure}

Various element abundances recorded in chondrites and planets exhibit a depletion trend correlated with their volatility, and its origin has been debated for the past several decades \citep{Anders1964, Palme1988, Palme1993, Bland2005, Hubbard2014}. The elements with condensation temperatures between that of silicon and sulfur in particular display a clear depletion trend, but this trend does not extend to elements more refractory than Si, including Al and Ca. Figure~\ref{fig_me} shows the Si-relative abundance of major elements for various types of chondrites, and a volatility-based depletion trend (i.e., higher abundance for more refractory elements) does not exist between highly and moderately refractory elements as seen in the bulk silicate Earth, ordinary chondrites, and enstatite chondrites. A simple volatility-based argument predicts that more refractory elements would have higher abundance because less refractory elements would have dissipated before condensing into solid. When the abundance of refractory elements is calculated using the approach of \cite{Cassen1996}, the CI-normalized ratio between Al and Si at 1-2~AU is predicted to be at least two. This ratio is much higher than what is observed in carbonaceous chondrites (up to 1.4) or in the bulk silicate Earth ($\sim$1.1). In ordinary and enstatite chondrites, Al and Ca even show depletion rather than enrichment. The lack of notable depletion trend between Al and Si indicates either that the highest temperature achieved in the disk must have been below the condensation temperature of Si (e.g., \citealp{Chick1997}), or that the argument based on volatility is not necessary valid. The former explanation is probably unrealistic because astrophysical models suggest that all the elements including refractory elements are likely to have evaporated around 1~AU when the mass accretion rate is larger than $10^{-7} M_\odot$ yr$^{-1}$ \citep{Bell1997}, which is a typical accretion rate for the disk at an early phase. 

Our dust settling model may bring a new insight into the compositional trend observed in chondrites. If we base ourselves solely on element volatility, it seems inevitable for them to be more enriched in Al and Ca compared to Mg and Si, if the initial temperature of the terrestrial region is higher than the condensation temperature of Si. As discussed in the previous section, however, corundum and gehlenite are stable at all height and do not form a condensation front. Thus, their grain growth is unlikely to be limited by the front, and they could possibly stick to each other and grow in size. Our model does not include grain growth, so the final size of such grains is uncertain, but it predicts a clear contrast between the behaviors of highly and moderately refractory elements. If Al and Ca-bearing grains become larger than cm-size, they tend to decouple from the gas and drift inward in a short time scale \citep{Weidenschilling1977}. Therefore, whereas dust bearing Mg, Si, and Fe is likely to be suppressed to $\mu$m-size above the condensation front, refractory grains probably drift inward, either concentrating in the inner region or falling into the Sun. This is consistent with the metoritic evidence that most of the early-formed refractory dust grains were removed before chondrite formation \citep{Brearley1988}. In Figure~\ref{fig_me}, we also show theoretical predictions based on our dust settling models, by calculating the composition of the vertical column, including both dust and gas, but excluding refractory grains below the condensation front. The range of depletion in Al and Ca in our prediction is greater than that in ordinary and enstatite chondrites, meaning that the observed depletion in these chondrites can be explained by changing the degree of removal of highly refractory grains. The degree of depletion depends on the continuation of condensation front, and the inner region is likely to be more depleted in highly refractory elements. %
The formation of spinel (MgAl$_2$O$_4$) below the front creates a small depletion of Mg as well. %
A small difference between the results from 1~and~2~AU originates in the formation of albite (NaAlSi$_3$O$_8$) above the condensation front, because Al condensed in the form of albite would not drift inward. %
The regions further from the Sun are colder and allow more albite to condense, resulting in weaker depletion of Al.

This has an important implication for the composition of the Earth and terrestrial planets at large. For example, Al and Ca are depleted compared to Mg in the BSE composition model of \cite{Lyubetskaya2007}, and this aspect is in direct conflict with the conventional expectation based on elemental volatility \citep{Palme2014}. Radioactive elements, including U and Th, are among highly refractory elements, and they are an important heat source for mantle convection. The formation of a condensation front could thus leave a long-lasting impact on the evolution for terrestrial planets as well. Predicting the BSE composition based on our modeling of vertical dust morion is not warranted because the formation of Earth-like planets should have involved the radial mixing of planetesimals, but the formation of condensation front, which can counteract the volatility-based elemental fractionation, has a potential to explain the apparently puzzling feature of the BSE composition, i.e., Al and Ca are more depleted with respect to Mg, but more enriched with respect to Si. %

Chondrites that formed further out where the condensation front did not form would not go through such enrichment nor depletion, and therefore, abundances close to CI chondrites are likely to be recorded.
How far the condensation front formed cannot sharply be constrained because of the uncertainty of disk parameters, but the boundary likely existed between 2~and~5~AU, which corresponds to the border between the source regions of carbonaceous and ordinary chondrites. This is in broad agreement with our model results for the extent of condensation front (Figure~\ref{fig_sum}).

\subsection{Outlook}
With our 1-D dust settling model, it is difficult to discuss the depletion trend in a more quantitative manner, because the effect of radial drift has yet to be considered \citep{Estrada2016}. Nevertheless, our model has shown the importance of the physical effect of condensation in the theory of planet formation, and we plan to extend the model to 2-D in the future. Although grain growth has a relatively small effect on vertical dust settling in the presence of a condensation front, it is likely to be important in radial models. %
Also, grain growth could cause variations in grain structure and surface composition, both of which could affect the opacity to some extent \citep{Semenov2003}. %
Turbulence would play a larger role in radial models, transporting a substantial amount of mass.

Some previous astrophysical models have already studied the effect of condensation on radial evolution. They showed that the effects of evaporation as dust grains drift inward could lead to an order-magnitude enhancement of volatile element abundance in the vicinity of condensation temperature \citep{Cuzzi2004, Ciesla2006}. However, most of the existing astrophysical models still treat condensation temperatures as if they were constant, although they may change by 100-200~K as pressure changes spatially and temporally throughout the disk. The enhancement of dust/gas ratio or the concentration of certain elements could also affect condensation temperatures significantly (\citealp{Ebel2000}; see also Appendix~A). Moreover, most of astrophysical models classify dust into only two classes, i.e., ice and silicate, although this treatment will lose the important information of condensing minerals, preventing quantitative comparison with meteoritic data.

Some radial models that tried to incorporate cosmochemistry (e.g., \citealp{Cassen1996, Ciesla2008}) have used an analytical disk model (e.g., \citealp{Chambers2009}). However, this ignores the effect of evolving dust composition on opacity. Opacity is sensitive to grain size, shape, and composition, and it should be calculated by honoring the disk environment \citep{Estrada2016}. In our approach, the effect of chemistry on dust migration is taken into account accurately by thermodynamic calculations, and the evolution of compositional variation within the disk is fed back to the dynamics of disk evolution through opacity calculation, making the calculation fully self-consistent. By comparing results from our future 2-D models with meteoritic data, we may be able to estimate some disk parameters such as initial density profile, viscosity, and grain size evolution, all of which are still highly uncertain. To extract such astrophysical constraints from cosmochemical data, it is of utmost importance to develop a new theory of protoplanetary disks, which is consistent with both physics and chemistry.

\section{Conclusion}
We constructed a vertical dust settling model by incorporating the effects of chemistry to the classical physical model of dust motion in a self-consistent manner. We showed that dust grains evaporate as they descend towards the hotter interior and cause the concentration of dust-composing elements. This leads to the formation of a condensation front.
Concentrated dust grains act as a high opacity layer, generating a large temperature increase, which hampers further dust settling by evaporation. The formation of a condensation front is ubiquitous at around 1~AU and could exist at as far as 4~AU depending on the choice of disk parameters. The existence of the front would significantly change the time scale of dust settling and also cause highly refractory grains (with Al or Ca) and moderately refractory grains (with Mg, Si, or Fe) to settle in different modes. Moderately refractory grains would form a condensation front and settle in a long time scale, but highly refractory grains would not experience evaporation and are likely to settle quickly towards the mid-plane. Highly refractory elements such as Al and Ca could end up in large grains and experience radial drifting. This could potentially explain the chemical diversity observed among the bulk silicate earth, ordinary chondrites, and carbonatious chondrites.

{\bf Acknowledgements.} This work was supported in part by the facilities and staff of the Yale University Faculty of Arts and Sciences High Performance Computing Center. We thank an anonymous reviewer for constructive comments on the earlier version of the manuscript.

\begin{table}
\caption{Chemical species considered in Gibbs free energy minimization}
\label{spec}
	\centering
	\begin{tabular}{c|l}
		\hline \hline
		phase & \multicolumn{1}{c}{species} \\
		\hline
Gas & Al, AlH, AlOH, Ca, Fe, H, H$_2$, H$_2$O, Mg, Na, NaOH, O, O$_2$, Si, SiO \\ \hline
Solid & corundum (Al$_2$O$_3$), melilite\footnotemark[1], olivine\footnotemark[2], orthopyroxene\footnotemark[3], metallic iron (Fe)\\
	& spinel (MgAl$_2$O$_4$), anorthite (CaAl$_2$Si$_2$O$_8$),\\
	& diopside (CaMgSi$_2$O$_6$), albite (NaAlSi$_3$O$_8$) \\ \hline
	\end{tabular}
	\footnotetext[1]{Solid solution of Ca$_2$Al$_2$SiO$_7$ (gehlenite) - Ca$_2$MgSi$_2$O$_6$ (akermanite)}
	\footnotetext[2]{Solid solution of Mg$_2$SiO$_4$ (forsterite) - Fe$_2$SiO$_4$ (fayalite)}
	\footnotetext[3]{Solid solution of MgSiO$_3$ (enstatite) - FeSiO$_3$ (ferrosilite)} 
\end{table}

\begin{table}
	\caption{Convergence test using the mid-disk temperature (in [K]) after 5$\times 10^4$ yr, with different numbers of grid cells. These cases are calculated at 1~AU with $\alpha = 10^{-3}$ and the initial disk mass of 0.21~$M_\odot$.}
	\label{conv}
	\centering
	\begin{tabular}{c|cccc}
		\hline
		No. of grids & 50         &  100      &  200    & 400  \\ \hline
		             & \multicolumn{4}{c}{T[K]} \\
		0.1~$\mu$m & 1385.9  &  1384.7 & 1385.2 & 1386.7 \\
		1~$\mu$m    & 1657.4  & 1653.3 & 1654.9 & 1655.2 \\
		10~$\mu$m  &  1389.3  & 1387.6 & 1388.2 & 1389.5 \\ \hline
	\end{tabular}
\end{table}


\appendix
\section{Gibbs Free Energy Minimization}
The total Gibbs free energy of the system that contains $m$~different species may be expressed as
\begin{equation} \label{n-Gibbs}
	G(n_1, n_2, \cdots, n_m) = \sum_{i=1}^{m} n_i \mu_i,
\end{equation}
where $n_i$ and $\mu_i$ denote the quantity and chemical potential of $i$-th species, respectively. For solid, the chemical potential is assumed to be independent of pressure, because the pressure effect is small in the low pressure condition. The chemical potential of gas is given by $\mu_i^0 + RT \ln (Pn_i/N)$ for gas, where $\mu_i^0$ is the chemical potential at a pressure of 1~atm and temperature $T$, and $N$ is the total number of moles of gaseous species. Equilibrium state is acquired by minimizing the Gibbs free energy under the constraint of mass balance equation:
\begin{equation} \label{mass_b1}
	\sum_{i=0}^{m} B_{ji} n_i = q_j \ \ \ (j=1,2,\cdots,d),
\end{equation}
where $d$ is the number of elements in the system, $B_{ji}$ is the number of atoms of element $j$ in $i$-th species, and $q_j$ is the number of moles of element $j$ in the system. Equation~(\ref{mass_b1}) could be written in a matrix form: $\bm{B^T n} = \bm{q}$. In the system consisted of H, O, Mg, and Al, for example, the corresponding row of Al$_2$O$_3$ in $\bm{B}$ will be $(0, 3, 0, 2)$. We use a non-linear conjugate gradient method to minimize Equation~(\ref{n-Gibbs}). We start with an initial composition of~$\bm{n^0}$ and incrementally update the composition along the direction calculated from the gradient at each iteration. The gradient of the free energy function, $(\nabla G)_i = \partial G / \partial n_i$, is given by $\mu_i^0$ for a solid species, and by $\mu_i^0 + RT \ln (P n_i/N)$ for a gas species. This gradient, however, does not satisfy the mass balance relationship, so we project the gradient to the null-space of $\bm{B^T}$ using the projection matrix, $\bm{P} = \bm{I} - \bm{B} (\bm{B^T B})^{-1} \bm{B^T}$, so that numbers of moles of the elements are conserved. By applying $\bm{B^T}$, mass conservation can be shown to hold: $\bm{B^T} \bm{P} \nabla G = 0$. The updated composition $\bm{n^{k+1}}$ is searched as a minimum along the direction of $\bm{p^k}$: $\bm{n^{k+1}} = \bm{n^{k}} + \alpha_k \bm{p^k}$, where $\alpha_k$ denotes an adjustable step length. The line search direction $\bm{p^k}$ is calculated using the Polak-Ribi\`ere method \citep{Polak1969}:
\begin{equation}
	\bm{p^k} = - \bm{g_k} + \frac{ \bm{g_{k}^T} (\bm{g_k} - \bm{g_{k-1}})}{ \bm{g_{k-1}^T} \bm{g_{k-1}}} \bm{p^{k-1}}\ \ \ (k\ge1),
\end{equation}
where $\bm{g_k}$ is the projected gradient, $\bm{P} \nabla G (\bm{n^k})$. For the initial search, $\bm{p^0}$ is set to $\bm{P} \nabla G(\bm{n^0})$. In order to calculate the step length~$\alpha_k$ that minimizes the free energy along the search direction~$\bm{p^k}$, the bisection method is adopted, by using the fact that the dot product of $\bm{p^k}$ and $\bm{P} \nabla G$ is zero at the minimum. Other search methods that directly compare the free energy are less reliable, because subtle changes in free energy caused by metal elements can be so small compared with the free energy of hydrogen that their effect cannot be represented by limited numerical precision.

\begin{figure}[ht!]
\epsscale{1}
\plotone{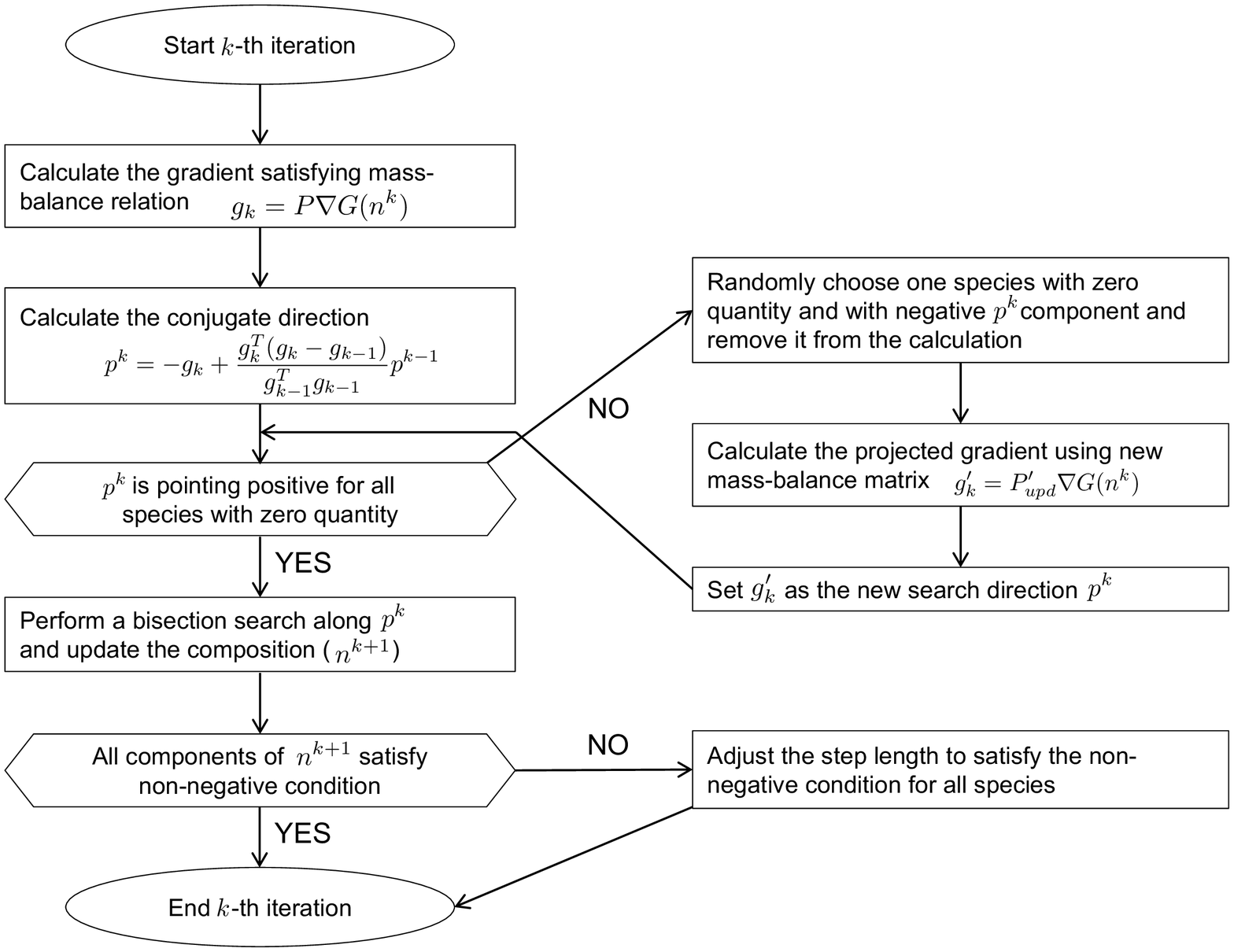}
\caption{Flowchart for one iteration in the new Gibbs free energy minimization scheme. This entire procedure is repeated as the composition is updated at each iteration. The iteration is terminated when the norm of the projected gradient becomes sufficiently small.}
\label{fig_flow}
\end{figure}

\begin{figure}[ht!]
\epsscale{0.95}
\plotone{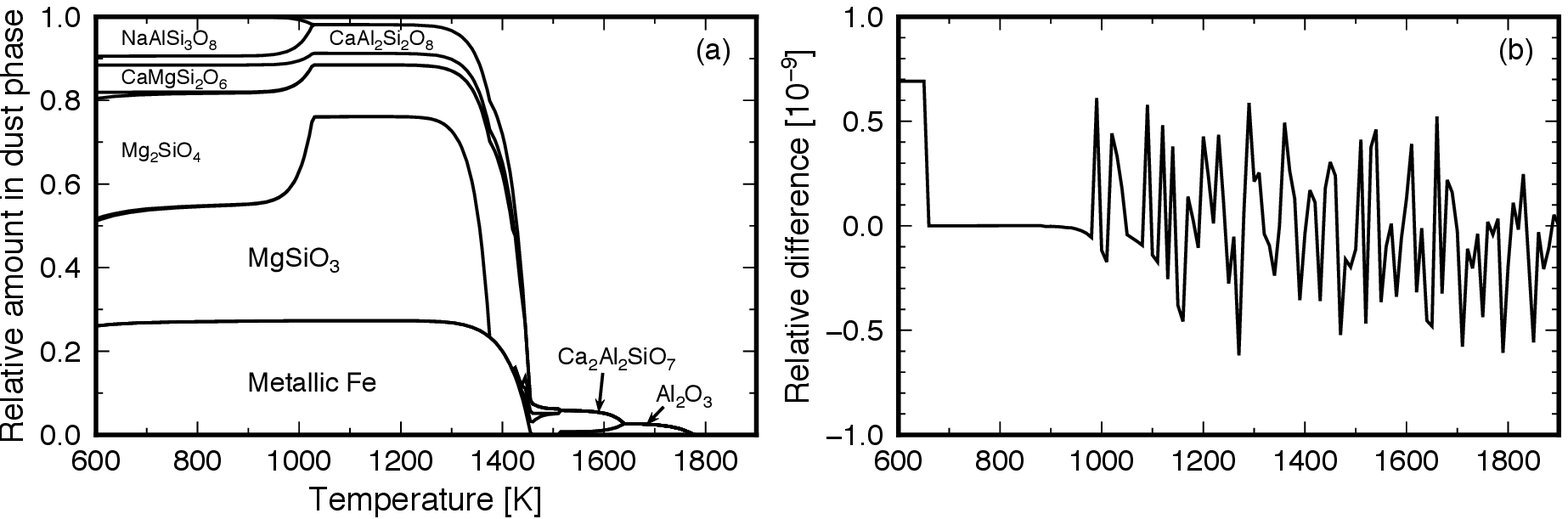}
\plotone{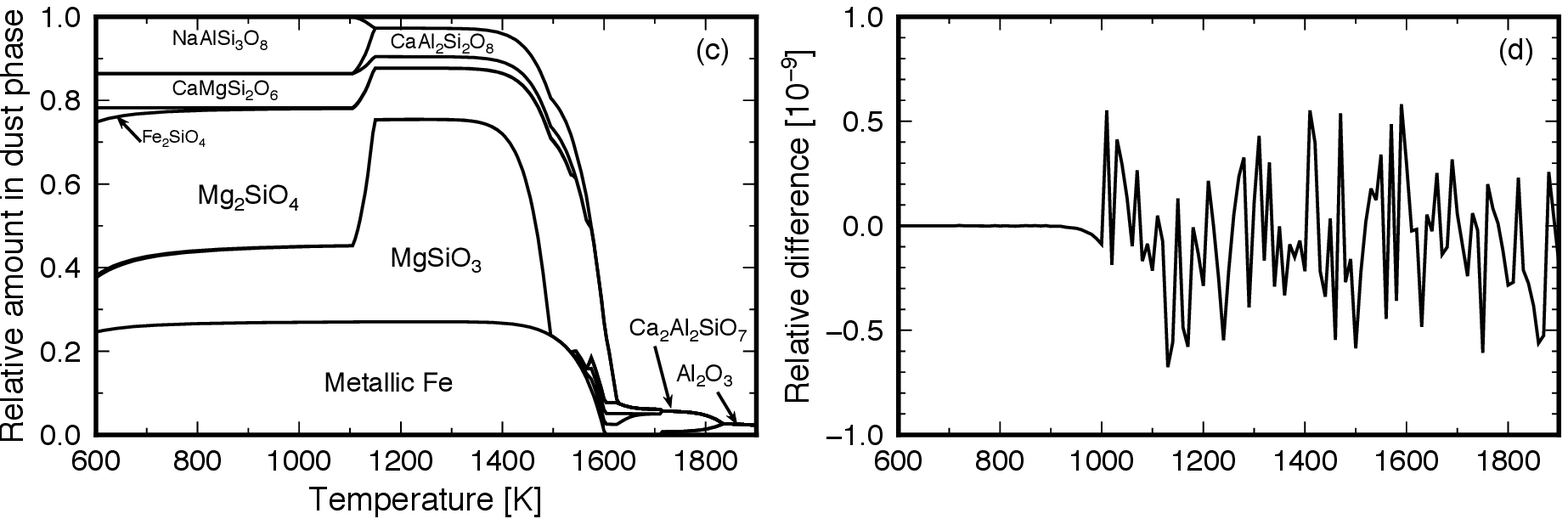}
\plotone{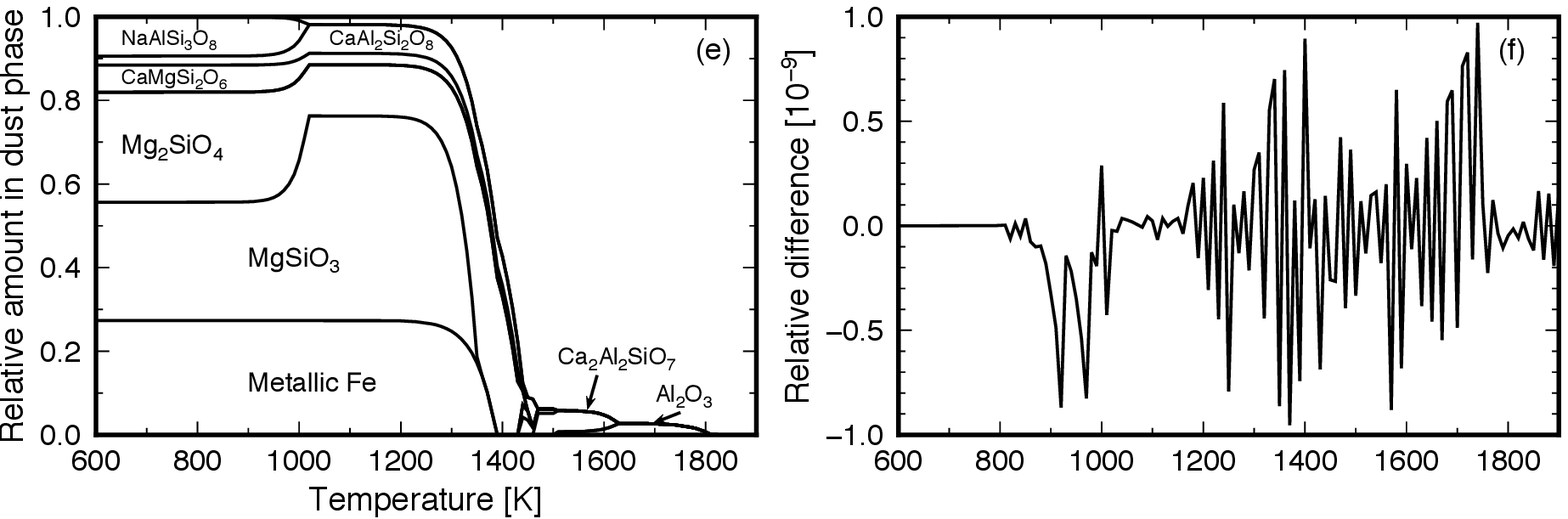}
\caption{(a) The Condensation sequences of major rock-forming phases at a total pressure of 10$^{-3}$~bar from a gas of solar composition. Relative molar amount of dust-composing elements, Na, Mg, Al, Si, Ca, and Fe, in the solid phase is plotted as a function of temperature. The solar abundance data taken from \cite{Lodders2003} and thermodynamic database adopted in the main section are used. (b) Relative difference in Gibbs free energy between HSC Chemistry, $G_{HSC}$, and our code, $G$, calculated as $(G-G_{HSC})/G_{HSC}$. The same composition with (a) but the thermodynamic database adopted in HSC are used for this benchmark calculation. (c and d) Same as (a) and (b), respectively, but with the dust-enriched composition containing 20~times more Na, Mg, Al, Si, Ca and Fe relative to the solar composition. \edy{(e and f) Same as (a) and (b), respectively, but with the 20~times dust-enriched composition at a lower total pressure of 10$^{-5}$~bar.}}
\label{fig_seq}
\end{figure}

Another constraint on minimizing the free energy is that the number of moles of every species should be non-negative. If the result of the line search returns a negative amount for a certain species, the step length~$\alpha_k$ is adjusted to satisfy the non-negative condition for all species. At a certain iteration, a number of moles $n_i$ could be zero for some $i$, but the $i$-th component of the search direction $\bm{p^k}$ could be negative, prohibiting further optimization. In this case, we remove the $i$-th species from our calculation for this step and re-optimize the number of moles. We shall, however, return $i$-th species into our calculation in the next iteration, because there is no guarantee that $n_i$ is zero at the global minimum. Similarly, when the amounts of several species are zero and their search directions are pointing towards negative, we should not remove all the species from the calculation at once.
Even if the $i$-th component of the original projected gradient is positive, the sign of the component may change by excluding some species. We should thus check whether each component of the projected gradient is positive or not every time we remove any of the species from our calculations. It is important to randomize the order of removing the species from the system, because whether the projected gradient is positive or negative depends on the set of species we choose to compose $\bm{B^T}$. It is possible that when we remove the species from our calculations in a certain order, the projected gradient could end up with a zero vector, but if we change the order of removing the species, we may obtain a non-zero gradient that points to a lower energy state, satisfying the non-negative molar amount constraint. Gas species may show a extremely small amount at global minimum, but its derivative might show a large value due to the steep nature of the derivative of log function even close to the minimum. When the gradient is large despite of the composition being close to the global minimum, it causes extra iterations to converge, so removing gas species with infinitesimal amount from the calculation at certain iterations is a way to make the calculation converge faster.
This series of procedures is continued until the norm of the projected gradient becomes sufficiently small. A flowchart for this procedure is shown in Figure~\ref{fig_flow}. 
The validity of our code is checked against HSC Chemistry (version 8.1) using the solar abundance data from \cite{Lodders2003}. For all of the temperature range considered, our code matches the result from HSC Chemistry within the order of numerical error (Figure~\ref{fig_seq}(c)).
The condensation sequences starting with solar abundance gas and dust-enriched gas are calculated using our method, and they are plotted in Figure~\ref{fig_seq}. The dust-enriched gas contains 20~times more Na, Mg, Al, Si, Ca and Fe relative to solar composition gas. It can be seen that the condensation temperatures of these elements are up to 100-200~K higher in the dust-enriched system.

\allauthors 
\listofchanges

\end{document}